\documentclass[11pt]{article}

\usepackage{times}

\usepackage{amssymb}  

\usepackage{epsfig} 
  \renewcommand*{\Pr}{\mathop{\mathrm{Prob}}}

  \def \qedbox{\hfill\vbox{\hrule\hbox{\vrule 
height1.3ex\hskip0.8ex\vrule}\hrule}}

  \newcommand{\goesto}{\rightarrow}
  \newtheorem{theorem}{Theorem}
  \newtheorem{definition}{Definition}
  \newtheorem{proposition}{Proposition}

\def\AND{\wedge}
\def\OR{\vee}

\def\goesto{\rightarrow}
\def\implies{\Rightarrow}

\newtheorem{lemma}{Lemma}[section]

\newtheorem{claim}{Claim}

\def\qed{\hfill$\Box$\newline\vspace{5mm}}

\begin{document}

\pagestyle{empty}

\title{Phase transitions and {\em all that}}
\author{Gabriel Istrate\thanks{e-mail: istrate@lanl.gov,
        NISAC, National Infrastructure Simulation Analysis Center, 
        Los Alamos National Laboratory,
        Mail Stop M 997, Los Alamos, NM 87545, U.S.A.}}
\date{}

\maketitle 
\thispagestyle{empty}
\section{Introduction}

Since the experimental paper of Cheeseman, Kanefsky and Taylor 
\cite{cheeseman-kanefsky-taylor}
{\em phase transitions in combinatorial problems} held the promise to 
shed 
light on the ``practical'' algorithmic complexity of combinatorial 
problems. However, the connection conjectured in 
\cite{cheeseman-kanefsky-taylor} was easily seen to be inaccurate. A much more realistic
possible connection has been highlighted by the results (based on 
experimental evidence and nonrigorous arguments from Statistical Mechanics) 
of Monasson et 
al. \cite{2+p:nature} (see also \cite{2+p:rsa}). These results 
supported 
the conjecture that it is {\em first-order phase transitions} that 
have algorithmic implications for the complexity of restricted 
classes of algorithms, including the important class of 
{\em Davis-Putnam-Longman-Loveland (DPLL) algorithms} \cite{beame:dp}. 

There exists, indeed, a nonrigorous argument supporting this 
conjecture:
phase transitions amount to nonanalytical behavior of a certain  
{\em order parameter}; the phase transition is {\em first order} if the 
order parameter is actually 
discontinuous. At least for random $k$-SAT \cite{monasson:zecchina} the 
order parameter suggested by Statistical Mechanics considerations has a 
purely combinatorial interpretation: it is the {\em backbone} of the 
formula, the set of 
literals that assume the same value in all optimal assignments. 
But intuitively one can relate (see e.g. 
the presentation of this argument by Achlioptas, Beame and Molloy 
\cite{achlioptas:beame:molloy:slides}) 
the size of the backbone to the complexity of DPLL algorithms, 
when run on random $k$-SAT instances slightly above the phase 
transition: All literals in the 
backbone require well-defined values in order to satisfy the 
formula. But a DPLL algorithm has very few ways to know what those 
``right'' values are. If w.h.p. the backbone of formulas above the 
transition contains a positive 
fraction of the literals that is bounded away from zero as we approach 
the transition (which happens in a case of a first-order phase 
transition) then, 
intuitively, DPLL will misassign a variable having $\Omega(n)$ height 
in the 
tree representing the behavior of the algorithm, and 
will be forced to backtrack on the given variable. 
{\em The conclusion of this intuitive argument is that a 
first-order phase transition implies a $2^{\Omega(n)}$ lower bound for 
the 
running time of any DPLL algorithm, valid with high probability for 
random instances located slightly above the transition.  }

While previous rigorous results 
\cite{2+psat:ralcom97,scaling:window:2sat,achlioptas:beame:molloy}, supported these intuitions, to date, the 
extent of a connection between first-order phase transitions and 
algorithmic complexity was unclear. 

{\bf The goals of this paper are
\begin{enumerate}
\item  To remedy this, and formally establish a connection between 
first-order phase 
transitions and the resolution complexity of random satisfiability 
problems, and
\item To take steps towards obtaining a complete classification of the 
order of phase transition in generalized satisfiability problems. 
\end{enumerate} 
}  

To accomplish these goals 

\begin{enumerate} 
\item we obtain (Theorem~\ref{dichotomy:threshold}) a complete 
characterization of sharp/coarse thresholds in the random generalized 
satisfiability model due to Molloy \cite{molloy-stoc2002}. 
``Phyisical'' arguments (see discussion below) imply that it makes no 
sense to study the order of the phase transition unless the problem 
has a sharp threshold. 
\item we rigorously prove (Theorem~\ref{3sat:first-order}) that random 
3-SAT has a first-order phase transition. We extend this result in 
several ways: first (Theorem~\ref{2+p-sat:first-order}) to 
random $(2+p)$-satisfiability, the original problem from 
\cite{2+p:nature}, obtaining further theoretical support to the 
heuristic results of \cite{2+p:nature}. Second we give a sufficient 
condition (Theorem~\ref{sufficient:first-order}) for the existence of 
a first-order phase transition. We then show 
(Theorem~\ref{implicates:first-order}) that all problems whose 
constraints have no implicates of size at most two satisfy this 
condition.\item we show that in all the cases where  
we can prove the 
existence of a first-order phase transition, such problems have   
a $2^{\Omega(n)}$ lower bound on their resolution complexity 
(and hence the complexity of DPLL algorithms as well \cite{beame:dp}). 
Indeed, the two phenomena ($2^{\Omega(n)}$ resolution complexity and 
the existence of a first-order phase transition) have 
common causes. 
\item in contrast, we show (Theorem~\ref{second:order}) that, for {\em 
any generalized 
satisfiability problem},  a second-order phase transition implies, for 
every $\alpha >0$, a $O(2^{\alpha \cdot n})$ upper 
bound on the resolution complexity of their random instances (in the 
region where most formulas are unsatisfiable).  
\end{enumerate} 

\section{Preliminaries} 

Throughout the paper we will assume familiarity with the general 
concepts of phase transitions in combinatorial problems (see e.g. 
\cite{martin:monasson:zecchina}), 
random structures \cite{bol:b:random-graphs}, proof complexity 
\cite{beame:proof:survey}. Some papers whose concepts and methods we use in 
detail (and we assume greater familiarity with) 
include \cite{friedgut:k:sat}, \cite{chvatal:szemeredi:resolution},  
\cite{ben-sasson:resolution:width}. 

Consider a monotonically increasing problem $A=(A_{n})$, 
 under the constant probability model $\Gamma(n,p)$, that independently 
sets to 1 with probability $p$ 
each bit of the random string. As usual, for 
$\epsilon >0$ let $p_{\epsilon}= p_{\epsilon}(n)$ define the canonical
probability such that $\Pr_{x \in \Gamma(n,p_{\epsilon}(n))}[x \in
A]= \epsilon$. 

The probability that a random sample $x$ satisfies property $A$ (i.e. 
$x\in A$) is a monotonically increasing function of $p$. {\em Sharp 
thresholds} are those for which this function has a ``sudden jump'' from 
value 0 
to 1:  

\begin{definition} \label{sharp}
Problem $A$ has a {\em sharp threshold} iff for every $0<\epsilon < 
1/2$, we have $\lim_{n\goesto \infty} \frac{p_{1-\epsilon}(n)- 
p_{\epsilon}(n)}{p_{1/2}(n)} = 0$. 
$A$ has {\em a coarse threshold} if for some $\epsilon > 0$ it holds 
that
$\underline{\lim}_{n\goesto \infty} \frac{p_{1-\epsilon}(n)- 
p_{\epsilon}(n)}{p_{1/2}(n)} > 0$. 
\end{definition} 

For satisfiability problems (whose complements are monotonically 
increasing) the constant probability model amounts to adding every constraint 
(among those 
allowed by the syntactic specification of the model) to the random 
formula independently 
with probability $p$. Related definitions can be given for the other 
two models for 
generating random structures, 
the {\em counting model} and {\em the multiset model} 
\cite{bol:b:random-graphs}. Under reasonable 
conditions \cite{bol:b:random-graphs} 
these models are equivalent, and we will liberally switch 
between them. In particular, for 
satisfiability problem $A$, and an instance $\Phi$ of $A$, 
$c_{A}(\Phi)$ will denote its {\em constraint density}, the ratio 
between 
the number of clauses and the number of variables of $\Phi$. To specify 
the random model in this 
latter cases we have to specify the constraint density as a function of 
$n$, the number of variables. 
We will use  $c_{A}$ to denote the value of the constraint density 
$c_{A}(\Phi)$ (in the counting/multiset 
models) corresponding to taking $p=p_{1/2}$ in the constant probability 
model. 
$c_{A}$ is a function on $n$ that is believed to tend to a constant 
limit as $n\goesto \infty$. However, Friedgut's proof 
\cite{friedgut:k:sat} of a sharp threshold in $k$-SAT (and our results) leave this issue 
open.

The original investigation of the order of the phase transition in 
$k$-SAT used an order parameter called {\em the backbone}.  
Bollob\'{a}s et al. \cite{scaling:window:2sat} have investigated the 
order of the phase transition in 2-SAT under a different order parameter, 
a ``monotonic version'' of the 
backbone called {\em the spine}. 

\begin{equation}\label{spine:initial} 
Spine(\Phi) = \{ x\in Lit | (\exists) \Xi \subseteq \Phi, \Xi \in SAT,
\Xi \AND \{\overline{x}\}\in \overline{SAT}\}. 
\end{equation}

They showed that random 2-SAT has a continuous (second-order) phase
transition: the size of the spine, normalized by dividing it by the 
number of 
variables, approaches zero ( 
as $n\goesto \infty$) for $c<c_{2-SAT}=1$, and is continuous 
at $c=c_{2-SAT}$. By contrast, nonrigorous 
arguments from Statistical Mechanics \cite{monasson:zecchina} imply the 
fact  
that for $3-SAT$ the spine jumps discontinuously from zero to 
positive values at the transition point $c=c_{3-SAT}$ 
(a first-order phase transition).  

It is easy to see that the intuition concerning the connection between 
the 
complexity of DPLL algorithms and the size of the backbone (discussed 
briefly in the introduction) extends to the spine as well. In this 
paper 
whenever we will discuss the order of a phase transition we will do it  
with respect to this latter order parameter.  

We would like to obtain a complete classification of 
the order of the phase transition in random satisfiability 
problems. 
A preliminary problem we have to deal with is characterizing those 
problems that 
have a sharp threshold: indeed, Physics considerations require that, 
in order that the study of the (order of the) phase transition to be 
meaningful, the order parameter (in the case of $k$-SAT the spine) has 
to be, w.h.p.,  concentrated around its expected value 
(in Physics parlance it is a {\em self averaging quantity}), and it is 
zero up to a certain value of the control parameter 
(in our case constraint density $c$) and positive above it. 
In the case of $k$-SAT these conditions imply the fact that 
$k$-SAT has a sharp threshold. The argument (a ``folklore'' one) 
can be formally 
expressed by the following 

\begin{lemma}\label{spine-threshold} Let $c$ be an arbitrary {\em 
constant} value for the constraint 
density function.  
\begin{enumerate}
\item If $c< \underline{lim}_{n\goesto \infty} c_{k-SAT}(n)$ then 
$\lim_{n \goesto \infty} \frac{|Spine(\Phi)|}{n} =0$. 
\item If for some $c$ there exists $\delta > 0$ such that w.h.p. (as 
$n\goesto \infty$) $\frac{|Spine(\Phi)|}{n} > \delta$ then 
$\lim_{n\goesto \infty}
 \Pr[\Phi \in SAT]= 0$, that is $c> \overline{lim}_{n\goesto \infty} 
c_{k-SAT}(n)$. 
\end{enumerate}
\end{lemma}

The argument is generic enough to extend to {\em all} constraint 
satisfaction problems. So a necessary condition for the study of 
the phase transition to be meaningful is that the problem have a 
sharp threshold.     

\section{Coarse and sharp thresholds of random generalized 
satisfiability 
problems}

In this section we obtain a complete classification of thresholds of 
random satisfiability problems, under 
 Molloy's recent model of random constraint satisfaction problems from 
\cite{molloy-stoc2002} (specialized to satisfiability problems, i.e. 
problems with domain $\{0,1\}$).  This affirmatively solves 
an open problem raised in \cite{creignou:daude:sat2002}.

\begin{definition}\label{model} 
Consider the set of all $2^{2^{k}}-1$ potential nonempty binary 
constraints on $k$ variables 
$X_{1}, \ldots, X_{k}$. We specify a probability distribution ${\cal 
P}$ which selects a single 
random constraint, and let ${\cal C}= supp({\cal P})$ be the set of 
constraints on which ${\cal P}$ 
assigns positive probability. 

A random formula from $SAT_{n,M}({\cal P})$ is specified by the 
following procedure: 
\begin{itemize} 
\item $n$ is the number of variables. 
\item $M$ is the number of clauses, chosen by the following procedure: 
first select, uniformly at 
random and with repetition $m$ hyperedges of the uniform hypergraph on 
$n$ variables. 
\item for each hyperedge choose a random ordering of the variables 
involved. Choose a random 
constraint according to ${\cal P}$ and apply it on the list of 
(ordered) variables. 
\end{itemize} 
$SAT({\cal C})$ refers to the random model corresponding to ${\cal P}$ 
being the uniform distribution on ${\cal C}$. 
\end{definition} 

It turns out we face a technical difficulty when studying sharp and 
coarse thresholds in Molloy's model; it cannot  be directly mapped onto 
the 
constant probability model for which the notion of a sharp threshold in 
Definition~\ref{sharp} works. The definition of a sharp threshold we 
need 
to employ is the one from \cite{molloy-stoc2002} 

\begin{definition}\label{sharp:2} 
$SAT({\cal P})$ is said to have {\em a sharp threshold of 
satisfiability} if there exists a function $c(n)$ bounded away from 0 such that, for 
any $\epsilon >0$ if $M<(c(n)-\epsilon) n$
then $SAT_{n,M}({\cal P})$ is a.s. satisfiable and if 
$M>(c(n)+\epsilon) n$ then $SAT_{n,M}({\cal P})$ is a.s. unsatisfiable. On the other 
hand, if there exist two functions $M_{1}(n), M_{2}(n)$ 
such that $M_{1}(n)/M_{2}(n)$ is bounded away from zero, and the 
satisfaction probability of random instances from $SAT_{n,M_{1}}({\cal P})$, 
$SAT_{n,M_{2}}({\cal P})$ is bounded away from 
both 0 and 1 then $SAT({\cal P})$ is said to have {\em a coarse 
threshold.} 
\end{definition} 

However, just as in \cite{molloy-stoc2002} (where this was done in the 
case when ${\cal P}$ is the uniform distribution), for $k\geq 3$ 
one can map Molloy's model onto a modified version of the constant 
probability model, defined as follows: 
Let $p_{1}, \ldots p_{r}$ be positive numbers between zero and 1. A 
random sample $x$ from the model $\Gamma_{p_{1},\ldots p_{r}}(n,p)$ is 
obtained in the 
following way: divide the bits of $x$ into $r$ equal groups. Set each 
of the bits in the $i$'th group to 1 independently with probability 
$p\cdot p_{i}$. For this model the definitions of $p_{\epsilon}$ and sharp/coarse threshold from Definition~\ref{sharp} carry over, and are equivalent to 
those from Definition~\ref{sharp:2}. 

Indeed, let $r$ be the cardinality of the support of distribution ${\cal P}$, 
and $p_{1}, \ldots, p_{r}$ be the associated positive 
probabilities. 

In its general setting Molloy's model is specified as follows: divide 
the potential constraints into 
groups of $rk!$ constraints, corresponding to all possible applications 
of the $r$ constraint templates on a fixed set of $k$ variables. 
For each such group, independently with probability $p$, we make the 
decision to include at least one of the constraints in the group 
with probability $p=\frac{M}{rk!{{n}\choose {k}}}$ (going from $M$ 
clauses to including each potential edge independently with probability $p$ 
can be done just as in the uniform case from \cite{molloy-stoc2002}). 

Each realization of  constraint constraint template $i$ is chosen with 
probability probability $p_{i}/k!$. Denote this model by $M(n,p,p_{1}, 
\ldots, p_{r})$. 

Defining $f(x) = [(1+x pp_{1}/k!)\cdot (1+xpp_{2}/k!) \cdot \ldots 
\cdot (1+xpp_{r})/k!]^{k!}- x$ we have $f(1)>0 $ and, since (by a simple 
calculus argument) 
the minimum of $f(x)$ over the choices of $p_{i}\geq 0$, $\sum p_{i}=1$  
 is obtained when one of them 
is 1 and the others are zero,  
\[
f(\frac{1}{1-p})\geq [1+\frac{p}{k!(1-p)}]^{k!}-\frac{1}{1-p}\geq 0.  
\]

Let $\alpha= \alpha(n) >0$  be the smallest solution of the equation 
$f(\alpha) =0$. Thus $\frac{1}{1-p} \leq  \alpha$.

Define, for $i=1,r$, 
\[
p^{\prime}_{i}= \frac{1/k! \cdot \alpha \cdot p_{i}}{1+1/k! \cdot 
\alpha pp_{i}}
\]

\begin{claim}
The following hold for any $p=\theta(n^{1-k})$:  
\begin{enumerate}
\item For every formula $\Phi$ such that no two constraints on the same 
set of variables appear in it, 
\[
\Pr_{M(n,p,p_{1}, \ldots, p_{r})}(\Phi)\geq 
\Pr_{\Gamma_{p^{\prime}_{1}, \ldots, p^{\prime}_{r}}(n,p)}[\Phi]. 
\]
Consequently 
\begin{eqnarray*}
\Pr_{M(n,p,p_{1}, \ldots, p_{r})}[SAT({\cal P})]\geq 
\Pr_{\Gamma_{p^{\prime}_{1}, \ldots, p^{\prime}_{r}}(n,p)}[SAT({\cal P})| \\ 
\mbox{ no two constraints on the same set of variables appear in }\Phi 
].
\end{eqnarray*}
\item On the other hand, there exists $f(n) = 1+o(1)$ such that for 
every formula $\Phi$ such that no two constraints on the same set of 
variables appear in it, 
\[
\Pr_{M(n,p,p_{1}, \ldots, p_{r})}(\Phi)\leq f(n) 
\Pr_{\Gamma_{p^{\prime}_{1}, \ldots, p^{\prime}_{r}}(n,p)}[\Phi]. 
\]
Consequently 
\begin{eqnarray*}
\Pr_{M(n,p,p_{1}, \ldots, p_{r})}[SAT({\cal P})]\leq 
(1+o(1))\Pr_{\Gamma_{p^{\prime}_{1}, \ldots, p^{\prime}_{r}}(n,p)}[SAT({\cal P})| \\ 
\mbox{ no two constraints on the same set of variables appear in }\Phi 
].
\end{eqnarray*}
\end{enumerate}
\end{claim}

Indeed, consider the set of constraints on a fixed set of given 
variables. The probability (under $\Gamma_{p^{\prime}_{1}, \ldots, 
p^{\prime}_{r}}$) that a given clause of type $i$ is 
included, and none of the others are is equal to 
$\frac{pp^{\prime}_{i}}{1- pp^{\prime}_{i}}\cdot 
[(1-pp^{\prime}_{1})\ldots (1-pp^{\prime}_{r})]^{k!} $. 
But
\[
\frac{pp^{\prime}_{i}}{1-pp^{\prime}_{i}}= \alpha pp_{i}/k!. 
\]

Also 
\[
1-pp^{\prime}_{i}= \frac{1}{1+\alpha pp_{i}/k!}, 
\]

so, by the definition of $\alpha$, 
 
\[
[(1-pp^{\prime}_{1})\ldots (1-pp^{\prime}_{r})]^{k!}= \frac{1}{\alpha}.
\]

This means that the probability that in a given set of constraints 
exactly one constraint (of type $i$)  is chosen is equal to $pp_{i}/k!$, 
the same as in model $M$. On the other 
hand the probability that {\em no} constraint is chosen is equal to  
$[(1-pp^{\prime}_{1})\ldots (1-pp^{\prime}_{r})]^{k!}= \frac{1}{\alpha}$. 
But the same probability in model $M$ 
is $1-p$, and we know that $1-p \geq \frac{1}{\alpha}$. In both model 
decisions on different sets of $k$ variables are independent. The 
conclusion is that $M$ assigns a larger probability 
than $\Gamma_{p^{\prime}_{1}, \ldots, p^{\prime}_{r}}$ to any sample 
$x$ to which it assigns positive probability. Point (1) follows. 

Point (2) has a similar proof: by calculus the maximum value of $f(x)$ 
is obtained when the $p_{i}$'s are equal, so 
\[
0 = f(\alpha) \leq (1+\frac{\alpha p}{rk!})^{rk!} - \alpha \leq 
e^{\alpha p} - \alpha. 
\]

Since $p=o(1)$, for large enough $n$ $e^{\alpha p} \leq 1+ \alpha p + 
(\alpha p)^{2}$, so $(\alpha p)^{2}+ \alpha (p-1) +1 > 0$, in other 
words
\[
\alpha (1-p) \leq 1+ (\alpha p)^{2}.
\]

But the ratio of the probabilities associated to any given $\Phi$ by 
$M$ and $\Gamma_{p^{\prime}_{1}, \ldots, p^{\prime}_{r}}(n,p)$ verifies 
\[
\frac{ \Pr_{M(n,p,p_{1}, \ldots, 
p_{r})}(\Phi)}{\Pr_{\Gamma_{p^{\prime}_{1}, \ldots, p^{\prime}_{r}}(n,p)}[\Phi]}\leq [\alpha (1-p)]^{rk! 
{{n}\choose {k}}}\leq (1+ (\alpha p)^{2})^{rk! {{n}\choose {k}}}. 
\]
Since $p=\theta(n^{1-k})$ and $k\geq 3$ the right-hand side is a 
function of $n$ that is $1+o(1)$. 

To prove the result it is enough to observe that for $k\geq 3$ the 
expected number of times a random formula in $\Gamma_{p^{\prime}_{1}, 
\ldots, p^{\prime}_{r}}(n,p)$ contains two different 
clauses on the same set of variables is $o(1)$, since that will imply 
that the satisfaction probabilities in the two models are related by a 
$1-o(1)$ factor. 
Indeed, this number is 
\[
{{n}\choose {k}} \cdot [\sum_{\alpha,\beta} 
\frac{pp^{\prime}_{\alpha}pp^{\prime}_{\beta}}{(k!)^{2}}], 
\]

where indices $\alpha,\beta$ span the set of different pairs of clauses 
from a group. 
Since each group is finite (contains $rk!$ clauses) and 
$p=\theta(n^{1-k})$, this expected value is $\theta(n^{2-k})$, which is $o(1)$ for 
$k\geq 3$. 
\qed

\begin{definition} 
Constraint $C_{2}$ is {\em an implicate of $C_{1}$}
 iff every satisfying assignment for $C_{1}$ satisfies $C_{2}$.  
\end{definition}

\begin{definition} 
A boolean constraint $C$ {\em strongly depends on a literal} if 
it has an unit clause as an implicate. 
\end{definition} 

\begin{definition} 
A boolean constraint $C$ {\em strongly depends on a 2-XOR relation} if 
$\exists i,j\in \overline{1,k}$ 
such that constraint ``$x_{i}\neq x_{j}$'' is an implicate of $C$. 
\end{definition} 

Our result is: 

\begin{theorem}\label{dichotomy:threshold} 
Consider a generalized satisfiability problem $SAT({\cal P})$ (that is 
not 
trivially satisfiable by the ``all zeros'' or ``all ones'' 
assignment). Let ${\cal C}= supp({\cal P})$.   
\begin{enumerate}
\item if some constraint in ${\cal C}$ strongly depends on one 
component then $SAT({\cal P})$ has a coarse threshold. 
\item if some constraint in ${\cal C}$ strongly depends on a 
2XOR-relation then $SAT({\cal P})$ has a coarse threshold. 
\item in all other cases $SAT({\cal P})$ has a sharp threshold. 
\end{enumerate}
\end{theorem}
 
\begin{proof}

\begin{enumerate}
\item 
Suppose some clause $C$ implies a unit clause. We claim that $SAT({\cal 
P})$ has a coarse threshold in the region where the expected number of 
clauses is $\theta(\sqrt{n})$. 

That the probability that such a formula is bounded away from zero in 
this region it is easy to see: consider a random formula with $c\sqrt n$ 
constraints, and let $H$ be the $k$-uniform formula hypergraph. 
The expected number of pairs of edges $C_{i}$, $C_{j}$ that have 
nonempty intersection is 
\[
{{c\cdot \sqrt n}\choose {2}}\cdot (1 - \frac{{{n-k}\choose {k} 
}}{{{n}\choose {k} }})\leq \frac{(ck)^{2}}{2}
\]

Therefore with constant positive probability (that depends on $c$), all 
vertices will have degree at most 1 in the hypergraph $H_{n}$, and the 
formula will be satisfiable. 

If, on the other hand, both positive and negative unit clauses 
are implicates of constraints in ${\cal P}$ then one can adapt the 
well-known lower bound on the probability of intersection of two 
random sets of size $\theta(\sqrt{n})$ to show that, with constant 
probability a random formula will contain two contradictory unit 
clauses as implicates, and be unsatisfiable. 

The proof is similar in the case when only one type of unit clauses 
(w.l.o.g 
assume it's the positive unit clauses) are implicates of constraints in 
${\cal C}$. Since $SAT({\cal C})$ is not trivial there exists a 
constraint $C_{1}\in {\cal C}$ with an implicate of the type 
$\overline{x_{1}}\OR \ldots \OR \overline{x_{b}}$, $b\geq 2$. We deal first with the 
case when there exists a constraint 
$C_{2}\neq C_{1}$ such that $C_{2}$ has an unit clause as implicate. 
Then it is easy to construct a formula $F$ consisting of $b$ copies of  
$C_{1}$and one copy of $C_{2}$ that implies the (unsatisfiable) formula 
$\{x_{1}, \ldots, x_{b},\overline{x_{1}}\OR \ldots \OR 
\overline{x_{b}}\}$. 
It is easy to see that the expected number of copies of $F$ in a random 
instance of $SAT({\cal P})$ with $\theta(\sqrt{n})$ clauses is 
constant, 
so the probability that the instance is unsatisfiable is bounded away 
from zero. 

Finally, in the case when the only constraint in ${\cal C}$ that has an 
unit 
implicate is $C_{1}$. In this case one can use a trick similar to that 
used 
in the last paragraph of subsection~\ref{all:together}: we use half of 
the 
random copies of $C_{1}$ to imply (random) unit clauses, and the other 
half 
to imply (random) copies of $\overline{x_{1}}\OR \ldots \OR 
\overline{x_{b}}$. This way we can produce, with constant probability, a copy of 
the formula $F$. 

\item 
Suppose now that $C$ does not fall in the first case but 
has a 2XOR implicate. In this case Creignou and 
Daud\'{e} have shown when ${\cal P}$ is the uniform distribution 
(and this extends directly to the case of a general 
probability distribution as well) that $p_{1/2}= \Omega(n^{1-k}))$ and  
the expected number of constraints is $\theta{n}$. Let $c_{SAT({\cal 
P})}\cdot n$ be the expected number of constraints corresponding to 
$p_{1/2}$. Then there exists $\delta > 0 $ such that, for every $n$, 
$c_{SAT({\cal P})}= c_{SAT({\cal P})}(n) > \delta$. 

Let us consider a random formula with  $c\cdot n$ of constraints. 
By the well known result on triangles in random graphs it follows that 
with positive probability one can use $C$ to create a ``contradictory
triangle''. Therefore it is easy to see that for every $c>0$ 
the satisfaction probability is bounded away from 1. It is easy to see 
than this statement, together with the fact that $c_{SAT({\cal P})}(n) 
> \delta$ together imply that $SAT({\cal P})$ has a coarse threshold.  
   
\item 
We will concentrate in 
the sequel on the last one. As discussed previously, for $k\geq 3$ 
Molloy's model can be mapped onto 
a version of the constraint probability model. In the case $k=2$ we can 
establish the existence of a sharp 
threshold in a direct manner, by the same method as the one used by 
Chv\'{a}tal and Reed for 2-SAT \cite{mickgetssome} 
(the complete proof of this case will be presented in the full 
version).  Indeed, by the first two points of the Theorem, and the assumption 
$k=2$ 
the only possible constraints in ${\cal P}$ can be constraints $x\OR 
y$, $\overline{x} \OR \overline{y}$, $\overline{x}\OR y$, $x \OR 
\overline{y}$, $x=y$, and 
the first two are always present.   

Let us now consider the case $k\geq 3$, using the modified version of 
the constant probability model. We note first that there exists a simple 
observation
 that allows us to reduce the problem to the case when ${\cal P}$ is 
the uniform probability: the Friedgut-Bourgain 
result on sharp/coarse threshold properties in monotone problems 
\cite{friedgut:k:sat} uses the following result, an easy consequence of the 
Mean Value Theorem: 
if a monotonic property $A$ does {\em not}  have a sharp threshold 
(under model $\Gamma(n.p)$) then there exists $p^{*}=p(n)$ and a constant 
$C>0$ such that (for 
infinitely many $n$) 

\begin{equation}\label{coarse}
p^{*}\cdot I(p^{*}) < C,
\end{equation}

where $I(p^{*})=\frac{d\mu_{p}(A)}{dp}|_{p=p^{*}}$. 

The same argument works when $A$ is considered under model 
$\Gamma_{p_{1},\ldots p_{r}}(n,p)$. Moreover, it is an easy consequence of Russo's 
Lemma for $\Gamma_{p_{1},\ldots p_{r}}(n,p)$
that if equation~\ref{coarse} holds for $p^{*}$ and some tuple 
${p_{1},\ldots p_{r}}$, then it also holds (with a different constant $C$) for  
$p^{*}$ and 
tuple $p_{1}= \ldots p_{r}=1/r$. In other words it is enough to obtain 
a contradiction to the assumption that $SAT({\cal P})$ did not have a 
sharp 
threshold in the case when ${\cal P}$ is the uniform probability, which 
is what we show next. 

\subsection{A base case} 

To prove the theorem in the uniform case we will first consider a 
``base case'' that is 
easier to  
explain, and will be of use in solving the general case: let $a,b$ be 
two integers 
(not necessarily equal), both greater or equal to 2. Let $S$ be a set 
consisting of two constraints $C_{1}, C_{2}$ of arity $a$,
respectively $b$, specified by 
$C_{1}= \overline{x_{1}}\OR \ldots \overline{x_{a}}$, $C_{2}= x_{1}\OR 
\ldots x_{b}$. One can represent $SAT(S)$ in the framework of 
Definition~\ref{model}
by ``simulating'' $C_{1}$, $C_{2}$ by suitable constraints of arity 
$\max\{a,b\}$. 

We first outline how to prove that $SAT(S)$ has a sharp threshold: 
we apply the Friedgut-Bourgain result \cite{friedgut:k:sat} and infer 
that if 
$SAT(S)$ did not have a sharp  threshold than, for some $\epsilon, 
\delta_{0}, K>0$ and some probability 
$p=p(n)\in [p_{\epsilon}, p_{1-\epsilon}]$ 

\begin{enumerate}
\item either $ 
\Pr_{p=p(n)} [\Phi \mbox{ contains some } F\in \overline{SAT}\mbox{ 
with }|F|\leq K] > \delta_{0}$, or  
\item there exists a fixed satisfiable formula $F_{0}$, $|F_{0}|\leq K$  
such that $
\Pr_{p=p(n)} [\Phi \AND F_{0} \in \overline{SAT}] - \Pr [ \Phi \in 
\overline{SAT}] > 
\delta_{0}$. 

\end{enumerate} 

One easy observation is that in the second alternative we can always 
assume 
that $F$ consists of a conjunction of 
unit clauses: if $F$ is satisfiable and satisfies (2), then so does 
the conjunction of unit clauses specifying 
one satisfying assignment of $F$. The first alternative is eliminated 
by a result (Proposition 4.6) from 
\cite{creignou:daude:sat2002}. 
The key to disproving the second alternative, in the case of $k$-SAT, 
is 
a geometric result, Lemma 5.7 in \cite{friedgut:k:sat}. 
We restate it here for completeness.

\begin{lemma}\label{friedgut}
For a sequence $A=(A_{n})$ of subsets of the $n$-dimensional hypercube, 
 $A\subseteq \{0,1\}^{n}$, define $A$ to be {\em 
$(d,m,\epsilon)$-coverable} if the probability for a union 
of a random choice of $d$ subcubes (hyperplanes) of codimension $m$ to 
cover $A$ is greater than $\epsilon$ for large enough $n$. 
 
Let $f(n)$ be any function that tends to infinity as $n\goesto \infty$. 
For fixed $k$, $d$, and $\epsilon$ any $A$ that is 
$(d,1,\epsilon)$-coverable is $(f(n),k, \epsilon)$-coverable. 
\end{lemma}

The connection with satisfiability can be explained as follows: the set 
$A$ 
in the application of the Lemma~\ref{friedgut} will (intuitively) refer 
to the set of satisfying assignments of random formula $\Phi$. 
Hyperplanes 
of codimension 1 are associated to unit clauses, more precisely to the 
set 
of assignments {\em forbidden} by a given unit clause. The fact that 
$A$ 
can be covered with probability $\epsilon$ by a union of $d$ random 
hyperplanes of codimension 1 parallels the fact that with probability 
$\epsilon$, 
$\Phi \AND F_{0}$ becomes unsatisfiable. This is what the result of 
Friedgut-Bourgain gives us (for $\epsilon = \delta_{0}$, under the 
assumption that $k$-SAT does not have a sharp 
threshold). Hyperplanes of 
codimension $k$ correspond to the set of assignments forbidden by a 
given $k$-clause, and the conclusion of the geometric lemma is that adding 
any small (but unbounded) number $f(n)$ of random 
$k$-CNF clauses to random formula $\Phi$ boosts the probability of {\em 
not} being satisfiable at least as much as the addition of the 
(constantly many) unit clauses in $F_{0}$.   

For small enough $f(n)$ this statement can be directly refuted, by 
concentration results for the binomial distribution (Lemma 5.6  in 
\cite{friedgut:k:sat}). A simpler 
and more general way to derive it is given as
Lemma 3.1 in \cite{achlioptas:friedgut:kcol}.  

The same outline works for the case we consider. To state the geometric 
result we need, however, to 
work with two types of hyperplanes: 

\begin{definition}
Let $H_{n}=\{0,1\}^{n}$ be the $n$-dimensional hypercube, and let 
$w_{i}$ denote the value of the $i$'th 
bit of element $w\in H_{n}$. 
A {\em positive hyperplane of codimension $d$} is a subset of points 
of  $H_{n}$ defined by a system of equations $
x_{i_{1}}= \ldots = x_{i_{d}}=1$, where the $x_{i}$'s are distinct 
variables. Negative hyperplanes have 
a similar definition. 
\end{definition}

Our version of the geometric Lemma is 

\begin{lemma}\label{geometric} 
For a sequence $A=(A_{n})$ of subsets of the $n$-dimensional hypercube, 
 $A_{n}\subseteq \{0,1\}^{n}$  define $A$ to be {\em $(n_{1}, d_{1}, 
d_{2},m_{1}, m_{2},\epsilon)$-coverable} if the probability of a union        
of a random choice of $d_{1}$ negative hyperplanes of codimension 
$m_{1}$ 
and $d_{2}$ positive hyperplanes of codimension $m_{2}$ to cover 
$A_{n}$ is at least $\epsilon$ if 
$n\geq n_{1}$. Let $f(n)$, $g(n)$ be any 
functions that tends to infinity as $n\goesto \infty$. For fixed 
$k_{1}$,$k_{2}$, $d$, and $0<\delta <\epsilon$, 
there exists $n_{2}$ that depends on $k_{1}, k_{2}, d,\epsilon,\Delta, 
n_{1}$ (but {\em not} $A$) such that for any 
$n\geq n_{2}$ any $A_{n}\subseteq \{0,1\}^{n}$ that is 
$(n_{1},d_{1},d_{2},1,1,\epsilon)$-coverable is 
$(n_{2},f(n),g(n),k_{1},k_{2}, \epsilon - \delta)$-coverable. 
\end{lemma}
We will in fact prove a stronger version of the Lemma: 

\begin{lemma}\label{geometric:2} 
For a sequence $A=(A_{n})$ of subsets of the $n$-dimensional hypercube, 
assume that  
\[
\Pr[A\subseteq H_{1}\cup \ldots H_{d}]\geq \epsilon
\]

for all $n\geq n_{1}$, 
where the $H_{i}$'s are random hyperplanes of codimension 1, 
$d_{1}$ of them negative, $d_{2}$ of them positive.

Let $f(n)$, $g(n)$ be any 
functions that tends to infinity as $n\goesto \infty$. For fixed 
$k_{1}$,$k_{2}$, $d$, and $\delta >0$, 
there exists $n_{*}=n(n_{1},d_{1},d_{2},k_{1}, k_{2}, 
\epsilon,\delta,f,g)$, (however it does {\em not} depend on $A$) 
such that for any 
$n\geq n_{*}$ and any $i$, $0\leq i\leq d$
\begin{equation}\label{conclusion}
Pr[A\subseteq P_{1}\cup \ldots \cup P_{\frac{if(n)}{d}}\cup N_{1}\ldots 
\cup N_{\frac{ig(n)}{d}}\cup H_{i+1}\cup 
\ldots \cup H_{d}]\geq 
Pr[A\subseteq H_{1}\cup \ldots \cup H_{d}]-\frac{i\delta}{d},  
\end{equation}

where the $N_{i}$'s are random negative hyperplanes of codimension 
$k_{1}$ and the 
$P_{i}$'s are random positive hyperplanes of codimension $k_{2}$. 
\end{lemma}

\begin{proof} 

It is easy to see that one can assume that $d|f(n)$, $d|g(n)$ (since it 
is enough to prove the lemma for 
$\overline{f}(n)=d\lfloor f(n)/d \rfloor$, $\overline{g}(n)=d\lfloor 
g(n)/d \rfloor$). 
 
We will prove the lemma by double induction on $d_{1}, d_{2}$. By 
symmetry we only need to consider two ``base cases:'' \\

{\bf Case 1: $d_{1}=0$, $d_{2}=1$} \\

In this case (and the dual, $d_{1}=1$, $d_{2}=0$) we can replace, for 
$i=1$, 
 equation~\ref{conclusion}
by the stronger: 
\begin{equation}\label{conclusion:d=1}
\Pr[A\subseteq P_{1}\cup \ldots \cup P_{f(n)}\cup N_{1}\ldots \cup 
N_{g(n)}]\geq 
1-\delta.  
\end{equation} 

The hypothesis implies that for $n\geq n_{1}$ 
there exist $n\cdot \epsilon$ positive hyperplanes of codimension $1$ 
such that 
\[
A_{n} \subseteq P^{(n)}_{1}\cap \ldots \cap P^{(n)}_{n\cdot \epsilon}. 
\]

We will assume, w.l.o.g., in what follows that $A$ is in fact {\em 
equal} to 
the right hand side. 
If $KP$ is a random positive hyperplane of 
codimension $k_{1}$ 
then 
\[
\Pr[A \not \subseteq KP] = 1-\frac{{{n\cdot \epsilon}\choose 
{k_{1}}}}{{{n}\choose {k_{1}}}}. 
\]  

Indeed, suppose KP is specified by the (random set of) equations  
$x^{(n)}_{1}= \ldots =x^{(n)}_{k_{1}}=1$. The condition that 
$A\subseteq KP$ is 
equivalent to 
\[
\{x^{(n)}_{1}, \ldots x^{(n)}_{k_{1}}\} \subseteq \{p^{(n)}_{1}, 
\ldots, p^{(n)}_{n\cdot \epsilon}\}, 
\]

where $\{p^{(n)}_{1}, \ldots, p^{(n)}_{n\cdot \epsilon}\}$ are the 
literals that 
specify the hyperplanes $P^{(n)}_{1}, \ldots, P^{(n)}_{n\cdot 
\epsilon}$.  

Thus the probability that $A$ is included in the union of $g(n)$ random 
positive hyperplanes $KP_{i}$ of codimension $k_{1}$ is at least $1 - 
\Pr[ (\forall i): A\not \subseteq KP_{i}]$, which is equal to 
\[
1 - (1-\frac{{{n\cdot \epsilon}\choose {k_{1}}}}{{{n}\choose 
{k_{1}}}})^{g(n)} \sim 1- (1-\epsilon^{k_{1}})^{g(n)}\goesto 1 \mbox{ 
as } 
n\goesto \infty.
\]

It follows that there exists $n_{*}=n(n_{1},d_{1},d_{2},k_{1}, k_{2}, 
\epsilon,\delta,f,g)$
such that for $n\geq n_{*}$ the 
left-hand side is larger than $1-\delta$. 

\vspace{5mm}
{\bf Case 2: $d_{1}+d_{2}>1$} \\

It is enough to prove that there exists 
$n_{i}=n(n_{1},d_{1},d_{2},k_{1}, k_{2}, \epsilon,\delta,f,g,i)$
such that~\ref{conclusion} holds, for a fixed value of $i$, $0\leq i 
\leq d$, when 
$n\geq n_{i}$. Then we can take $n_{*}=max\{n_{i}\}$. 

We prove this on induction on $i$. The claim is clearly true for $i=0$. 
Assume, therefore, that 
the claim is true up to $i$; we will prove it for  $i+1$. 

Denote for all $j$
\[
p_{j}= \Pr[A \subseteq P_{1}\cup \ldots \cup 
P_{\frac{(j-1)f(n)}{d}}\cup N_{1}\cup \ldots \cup 
N_{\frac{(j-1)g(n)}{d}}\cup H_{j}\cup \ldots \cup H_{d}]. 
\]

To accomplish that it is enough to show that 
\begin{equation}\label{inductive:step}
p_{i+1}\geq p_{i}-\delta, 
\end{equation} 
 
By the Bayes formula: 
 
\begin{eqnarray*}\label{bayes}
p_{i}=\Pr[A \subseteq P_{1}\cup \ldots \cup P_{\frac{(i-1)f(n)}{d}}\cup 
N_{1}\cup \ldots \cup 
N_{\frac{(i-1)g(n)}{d}}\cup H_{i}\cup \ldots \cup H_{d}]  =  \\
 =  \sum_{B} 
\ Pr[B\subseteq H_{i}| A\setminus (P_{1}\cup \ldots \cup 
P_{\frac{(i-1)f(n)}{d}}\cup N_{1}\cup \ldots \cup 
N_{\frac{(i-1)g(n)}{d}}) = B] \cdot \\
\cdot \Pr[A\setminus (P_{1}\cup \ldots \cup P_{\frac{(i-1)f(n)}{d}}\cup 
N_{1}\cup \ldots \cup 
N_{\frac{(i-1)g(n)}{d}}) = B]
\end{eqnarray*} 

Assume without loss of generality that $H_{i}$ is a positive 
hyperplane. 

Let $\gamma= \frac{\delta}{2d}$. Let 
\[
C_{\gamma}=\{B\subseteq \{0,1\}^{n}: Pr[B\subset P] > \gamma\}.
\]

Let $\alpha$ be the sum of those terms in~\ref{bayes} corresponding to 
sets $B\in C_{\gamma}$, 
and let $\beta$ be the sum corresponding to sets $B\in 
\overline{C_{\gamma}}$. 

From the definition of $C_{\gamma}$ it follows that 
\[
0\leq \beta \leq \gamma,  
\]
 therefore
\begin{eqnarray*}\label{inequality} 
\Pr[A\setminus (P_{1}\cup \ldots \cup P_{\frac{(i-1)f(n)}{d}}\cup 
N_{1}\cup \ldots \cup 
N_{\frac{(i-1)g(n)}{d}}) \in C_{\gamma}]\geq \\ 
\geq \frac{1}{\gamma}\cdot [\Pr[A \subseteq P_{1}\cup \ldots \cup 
P_{\frac{(i-1)f(n)}{d}}\cup N_{1}\cup \ldots \cup 
N_{\frac{(i-1)g(n)}{d}}\cup H_{i}\cup \ldots \cup H_{d}] -\gamma]= \\
= \frac{1}{\gamma}\cdot [p_{i}-\gamma].
\end{eqnarray*} 

On the other hand 

\begin{eqnarray*}\label{bayes:2}
p_{i+1}=\Pr[A \subseteq P_{1}\cup \ldots \cup P_{\frac{if(n)}{d}}\cup 
N_{1}\cup \ldots \cup 
N_{\frac{ig(n)}{d}}\cup H_{i+1}\cup \ldots \cup H_{d}]  =  \\
 =  \sum_{B} 
\ Pr[B\subseteq P_{\frac{(i-1)f(n)}{d}+1}\cup \ldots \cup 
P_{\frac{if(n)}{d}} 
\cup N_{\frac{(i-1)g(n)}{d}+1}\cup \ldots \cup N_{\frac{ig(n)}{d}}| \\
A\setminus (P_{1}\cup \ldots \cup P_{\frac{(i-1)f(n)}{d}}\cup N_{1}\cup 
\ldots \cup 
N_{\frac{(i-1)g(n)}{d}}) = B] \cdot \\
\cdot \Pr[A\setminus (P_{1}\cup \ldots \cup P_{\frac{(i-1)f(n)}{d}}\cup 
N_{1}\cup \ldots \cup 
N_{\frac{(i-1)g(n)}{d}}) = B]
\end{eqnarray*} 

Let $\overline{f}(n)=f(n)/d$, $\overline{g}(n)= g(n)/d$. Since the 
$N_{i}$'s, $P_{i}$'s 
are random hyperplanes, one can 
rewrite the previous recurrence as 

\begin{eqnarray*}\label{bayes:3}
p_{i+1}=\Pr[A \subseteq P_{1}\cup \ldots \cup P_{\frac{if(n)}{d}}\cup 
N_{1}\cup \ldots \cup 
N_{\frac{ig(n)}{d}}\cup H_{i+1}\cup \ldots \cup H_{d}]  =  \\
 =  \sum_{B} 
\Pr[B\subseteq P_{1}\cup \ldots \cup P_{\overline{f(n)}} 
\cup N_{1}\cup \ldots \cup N_{\overline{g(n)}}| \\
A\setminus (P_{1}\cup \ldots \cup P_{\frac{(i-1)f(n)}{d}}\cup N_{1}\cup 
\ldots \cup 
N_{\frac{(i-1)g(n)}{d}}) = B] \cdot \\
\cdot \Pr[A\setminus (P_{1}\cup \ldots \cup P_{\frac{(i-1)f(n)}{d}}\cup 
N_{1}\cup \ldots \cup 
N_{\frac{(i-1)g(n)}{d}}) = B]
\end{eqnarray*}

Since all terms are nonnegative, one can obtain a lower bound on the 
left-hand size of this 
latter terms by only considering those $B\in C_{\gamma}$. 

Applying the induction hypothesis from case one for all $B\in 
C_{\gamma}$ and 
$n\geq 
n_{i}=n_{*}(n_{1},0,1,k_{1},k_{2},\gamma,\delta,\overline{f},\overline{g})$,  we 
infer that for all such $B$, 
\[
\Pr[B\subseteq P_{1}\cup \ldots \cup P_{\overline{f(n)}} 
\cup N_{1}\cup \ldots \cup N_{\overline{g(n)}}]\geq (1-\gamma), 
\]
therefore 
\begin{eqnarray*}
p_{i+1}= \Pr[A \subseteq P_{1}\cup \ldots \cup P_{\frac{if(n)}{d}}\cup 
N_{1}\cup \ldots \cup 
N_{\frac{ig(n)}{d}}\cup H_{i+1}\cup \ldots \cup H_{d}]\geq \\
(1-\gamma)\cdot 
\sum_{B\in C_{\gamma}} Pr[A\setminus (P_{1}\cup \ldots \cup 
P_{\frac{(i-1)f(n)}{d}}\cup N_{1}\cup \ldots \cup 
N_{\frac{(i-1)g(n)}{d}}) = B] \\
= (1-\gamma) \cdot Pr[A\setminus (P_{1}\cup \ldots \cup 
P_{\frac{(i-1)f(n)}{d}}\cup N_{1}\cup \ldots \cup 
N_{\frac{(i-1)g(n)}{d}})\in C_{\gamma}] \\ 
\geq \frac{(1-\gamma)}{\gamma}\cdot [\Pr[A \subseteq P_{1}\cup \ldots 
\cup P_{\frac{(i-1)f(n)}{d}}\cup N_{1}\cup \ldots \cup 
N_{\frac{(i-1)g(n)}{d}}\cup H_{i}\cup \ldots \cup H_{d}] -\gamma] = \\
\frac{(1-\gamma)}{\gamma}\cdot [p_{i}-\gamma].  
\end{eqnarray*}

Since $\gamma \leq 1$, 
\[
p_{i+1}\geq (1-\gamma)\cdot (p_{i}-\gamma)= p_{i} - \gamma \cdot 
[1+p_{i}-\gamma]\geq p_{i}-2\gamma.  
\]

which is precisely equation~\ref{inductive:step} (that we wanted to 
prove). 

\end{proof}
\qed

\subsection{How to contradict Lemma~\ref{geometric}}

It is now easy to infer the fact that $SAT(S)$ has a sharp threshold, 
by 
using the previous lemma with $d_{1}= |F_{0}\cap Var|$, 
$d_{2}=|F_{0}|-d_{1}$, 
$k_{1}=b$, $k_{2}=a$, $\epsilon = \delta_{0}$, $\Delta = \delta_{0}/2$ 
and a refutation of the geometric lemma similar to the 
one for 3-SAT.  

The conclusion of the Geometric Lemma (similar to the one for $k$-SAT) 
is that, by adding any number $f(n)$ of copies of $x_{1}\OR \ldots \OR 
x_{a}$ 
and $g(n)$ copies of $\overline{x_{1}}\OR \ldots \OR x_{b}$ suffices to 
boost the unsatisfiability probability by a constant. 

However, Lemma 3.1 from \cite{achlioptas:friedgut:kcol} asserts that 
adding up to $o(\sqrt(n))$ random clauses is not enough to boost the 
unsatisfiability 
probability by more than $o(1)$. 

Because of the nature of the random model, adding $H(n)=o(\sqrt(n))$ 
random clauses insures that w.h.p. we have $\Theta(H(n))$ copies 
of each type of clause, as long as $H(n)$ grows faster than some power 
of $n$. Taking the number of such copies to be the functions 
$f(n)$, $g(n)$ contradicts the consequence of the Geometric Lemma.

Note that we do {\em not} make use of all the details of the random 
model (such as the precise number of copies of each clause in random 
instances at $p=p(n)$), but only that: 
\begin{itemize}
\item the expected number of copies of both $C_{1}$ and $C_{2}$ in a 
random formula at $p=p(n)$ is unbounded. This is enough to make the 
analog 
of Lemma 3.1 from \cite{achlioptas:friedgut:kcol} work.  
\item  the clauses are independent. 
\end{itemize}  

\subsection{Putting it all together} \label{all:together}

In the previous section we have proved a geometric lemma that is used 
to prove that the above-defined set $SAT(S)$ has a sharp threshold. 

Consider now a set ${\cal C}$ of constraints that satisfies the 
condition (3) 
of the theorem. Since $SAT({\cal C})$ is not trivially satisfied by the 
``all zero'' (all ones) assignment, there exist constraints $C_{1}$, 
$C_{2}$ in ${\cal C}$ and $a,b \geq 1$ such that $C_{1}\models x_{1}\OR 
\ldots x_{a}$, 
$C_{2}\models \overline{x_{1}}\OR \ldots \OR \overline{x_{b}}$. In fact 
$a,b \geq 2$, otherwise some constraint in ${\cal C}$ would strongly 
depend on one variable. 

Just as in the base case, condition (i) in the Friedgut-Bourgain result 
is eliminated by the result of Creignou and Daud\'{e}, and  the formula 
$F_{0}$ in condition 
(ii) can be assumed to consist of a conjunction of unit clauses.
Reflecting this fact, the geometric lemma needed for the
general case has the same hypothesis as 
the one of Lemma~\ref{geometric}: the set $A$ can be covered by a union 
of random hyperplanes of codimension 1. 
However, the covering desired in the conclusion no longer consists 
of hyperplanes, but of (general) sets of points in the hyperplane, 
corresponding to sets of assignments forbidden by a certain constraint 
$C$. 

The critical observation (easiest to make in the case when $C_{1}\neq 
C_{2}$ )
is that {\bf Lemma~\ref{geometric} implies the geometric result we need 
for this case}: 
since $C_{1}\models x_{1}\OR \ldots x_{a}$, the ``forbidden set'' 
associated to  $C_{1}$ contains the ``forbidden set'' associated to 
constraint 
$x_{1}\OR \ldots x_{a}$ in Lemma~\ref{geometric}, the hyperplane 
$x_{1}= \ldots = x_{a}=0$. A similar result holds for $C_{2}$ 
and the positive hyperplanes. 

Thus each ``covering set'' in a conclusion of the (general case of 
the ) geometric lemma contains a corresponding ``covering set'' from 
the conclusion of Lemma~\ref{geometric}. In other words {\bf the 
geometric lemma for $SAT({\cal C})$ follows from the geometric lemma for the 
base case by monotonicity}.  

All is left to show is that the analog of Lemma 3.1 from
\cite{achlioptas:friedgut:kcol} also 
works in this general case. We have previously observed that this 
amounts 
to showing that the expected number of copies of $C_{1}$, $C_{2}$ in 
a random formula is unbounded. But Proposition 3.5 in 
\cite{creignou:daude:sat2002} (slightly generalized to probability distributions ${\cal 
P}$ that are not uniform) and the details of the random model imply 
that in fact this number is linear. 

A simple modification of this argument holds when $C_{1}=C_{2}$. To see 
this, note that in the proof of the fact that 
Lemma~\ref{geometric} holds for {\em some} small enough (but unbounded) 
$f(n),g(n)$ is enough to obtain a contradiction. 
 
The expected number of copies of $C_{1}$ in a random formula is linear, 
denote it by $h(n)$. Dividing the set of such copies into two (random) 
sets of cardinality $h(n)/2$ yields 
infinitely many random copies of $C_{1}$ 
used to imply  $x_{1}\OR \ldots \OR x_{a}$ in the previous argument
and infinitely many  copies used to imply   
$\overline{x_{1}}\OR \ldots \OR \overline{x_{b}}$. We then apply the 
same strategy as in the first case. 

To summarize: the proof follows from the corresponding 
argument for the base case by monotonicity. It 
critically uses the fact that we are {\em not} in cases (i) or (ii) of 
the Theorem, since it is only under these conditions when the first 
alternative in the 
Friedgut-Bourgain argument can be eliminated.

\end{enumerate} 
\end{proof}
\qed

\section{3-SAT has a first-order phase transition}

\begin{theorem}\label{3sat:first-order}
$k$-SAT, $k\geq 3$ has a first-order phase transition. In other words
there exists  
$\eta > 0 $ such that  for every sequence $p = p(n)$
 we have 
\begin{equation}\label{jump} 
\lim_{n\goesto \infty} \Pr_{p=p(n)}[\Phi \in SAT] = 0 \implies
\lim_{n\goesto \infty}\Pr_{p=p(n)}[ \frac{|Spine(\Phi)|}{n}\geq \eta]= 
1. 
\end{equation}  
\end{theorem}

\begin{proof}

We start by giving a simple sufficient condition for a literal to
belong to the spine of the formula: 

\begin{claim}\label{spine:unsat}
Let $\Phi$ be a minimally unsatisfiable formula, and let $x$ be a
literal that appears in $\Phi$. Then $x\in Spine(\Phi)$. 
\end{claim}
\begin{proof}
Let $C$ be a clause that contains $x$. By the minimal unsatisfiability
of $\Phi$, $\Phi \setminus \{C\} \in SAT$. On the other hand $\Phi 
\setminus \{C\}
\AND \{x\} \in \overline{SAT}$, otherwise $\Phi$ would also be
satisfiable. 
Thus $x \in Spine(\Phi \setminus \{C\})$. 
\end{proof}
\qedbox

Thus, to show that 3-SAT has a first-order phase transition it is 
enough to show that a random unsatisfiable formula contains w.h.p. 
a minimally unsatisfiable subformula containing a linear number of 
literals. A way to accomplish this is by using the two ingredients of 
the Chv\'{a}tal-Szemer\'{e}di proof \cite{chvatal:szemeredi:resolution} 
that random 3-SAT has exponential
resolution size w.h.p. They are explicitly stated to make the 
argument self-contained:   

\begin{claim} 
There exists a constant $\delta>0$ such that for every constant $c>c_{3 
SAT}$ with high 
probability (as $n\goesto \infty$) a random formula $\Phi$ with $n$ 
variables and $cn$ 
clauses has no minimally unsatisfiable subformula of size less than 
$\delta \cdot n$.  
\end{claim}

\begin{claim}
There exists $\eta >0$ so that w.h.p. for every $c>c_{3-SAT}$ all 
subformulas 
of a random formula $\Phi$ having between $(\delta/2)\cdot n$ and 
$\delta\cdot n$
clauses contain at least $\eta\cdot n$ (pure) literals (corresponding 
to different variables).  
\end{claim}

The argument is now transparent: if $\Phi$ is unsatisfiable then
w.h.p. a minimally unsatisfiable subformula $\Xi$ of $\Phi$ has 
size at least $\delta n$. By the second claim, applied to an arbitrary 
subformula of $\Xi$ of size $(3\delta n)/4$, we infer that w.h.p. $\Xi$ 
contains at least many $\eta\cdot n$ different variables.  

\end{proof}

\section{First-order phase transitions and resolution complexity 
of  random generalized satisfiability problems} 

In this section we extend the previous result (and the connection 
between first-order phase transitions and resolution complexity) to 
other 
classes of satisfiability problems. Interestingly, we find that a 
condition 
Molloy investigated in \cite{molloy-stoc2002} is a sufficient condition 
for the 
existence of a first-order phase transition. 

It turns out that there are differences between the case of random 
$k$-SAT and the general case that force us to employ an 
alternative definition of the spine. The most obvious one is that 
formula (~\ref{spine:initial}) involves 
negations of variables, whereas Molloy's model does not. But it has 
more serious problems: consider for example {\em $k$-uniform hypergraph 
2-coloring}
specified as $SAT(\{C_{0}\})$, where $C_{0}(x_{1}, \ldots, x_{k})$ has 
the interpretation ``not all of $x_{1}, \ldots x_{k}$ are equal''. 
Because of the built-in symmetry to permuting colors 0 and 1,  the 
spine of {\em any} instance is empty (under 
definition~\ref{spine:initial}). 
Similar phenomena have appeared before (and forced a different 
definition of the backbone/spine) 
in  {\em k-coloring \cite{frozen:development}} (symmetry = permutation 
of colors) or {\em graph partition} \cite{graph-partition:transition} 
(symmetry = permutation of sides). 
There are other ways (to be detailed in the journal version of the 
paper) in which the original definition of the 
spine behaves differently in the general case than in that of random 
$k$-SAT. Our solution is to define the concept of spine of a 
random instance of a satisfiability problem $SAT(\cal P)$ in a slightly 
different way. The definition  is consistent with those in 
\cite{frozen:development}, \cite{graph-partition:transition}. 

\begin{definition}$
Spine(\Phi) = \{ x\in Var | (\exists) \Xi \subseteq \Phi, \Xi \in SAT, 
(\exists) C\in {\cal C}, x\in C,\mbox{ such that } 
\Xi \AND C  \in \overline{SAT}\}$. 

\end{definition} 

It is easy to see that, for $k$-CNF formulas whose (original) spine 
contains at 
least three literals a variable $x$ is in the (new version of the) 
spine if and only if either $x$ or $\overline{x}$ 
were present in the old version. In particular the new definition does 
not change the order of the 
phase transition of random $k$-SAT. Moreover the proof of 
Claim~\ref{spine:unsat} carries over to the general case.  
The {\em resolution complexity} of an instance $\Phi$ of $SAT({\cal 
P})$ is defined as the resolution complexity of the formula obtained by 
converting each constraint of $\Phi$ to CNF-form.

\begin{definition} Let ${\cal P}$ be such that $SAT({\cal P})$ has a 
sharp 
threshold.
Problem $SAT({\cal P})$ has a {\em first-order phase transition} if
there exists  
$\eta > 0 \mbox{ such that  for every sequence }  p = p(n)$
 we have 
\begin{equation}\label{jump:general} 
\lim_{n\goesto \infty} \Pr_{p=p(n)}[\Phi \in SAT] = 0 \implies
\lim_{n\goesto \infty}\Pr_{p=p(n)}[ \frac{|Spine(\Phi)|}{n}\geq \eta]= 
1. 
\end{equation}  
If, on the other hand, for every $epsilon >0$ there exists $p^{\epsilon}(n)$ 
with 
\begin{equation}\label{jump:continuous} 
\lim_{n\goesto \infty} \Pr_{p=p^{\epsilon}(n)}[\Phi \in SAT] = 0 \mbox{ and }
\lim_{n\goesto \infty}\Pr_{p=p(n)}[ \frac{|Spine(\Phi)|}{n}\geq \epsilon]= 
0 
\end{equation} 
we say that $SAT({\cal P})$ has a {\em second-order phase 
transition}\footnote{strictly speaking the order of the 
phase transition is {\em at least two}.}.   
\end{definition} 

A first observation is that a second-order phase transition 
has computational implications: 

\begin{theorem}\label{second:order} 
Let ${\cal P}$ be such $SAT({\cal P})$ has a second-order phase 
transition. Then for every constant
$c>\overline{lim}_{n\goesto \infty} c_{SAT({\cal P})}(n)$, and {\em 
every $\alpha>0$}, random formulas of constraint density $c$ 
have w.h.p. resolution complexity $O(2^{k\cdot \alpha \cdot n})$.  
\end{theorem} 

\begin{proof} 

By the analog of Claim~\ref{spine:unsat} for the general case, if 
$SAT({\cal P})$ has a second-order phase transition) 
then for every $c>c_{SAT({\cal P})}$ and for every $\alpha>0$, 
minimally unsatisfiable subformulas of 
a random formula $\Phi$ with constraint density $c$ have w.h.p. size at 
most $\alpha \cdot n$.  
Consider 
the backtrack tree of the natural DPLL algorithm (that tries to 
satisfies clauses one at a time) 
on such a minimally unsatisfiable subformula $F$. By the usual 
correspondence between DPLL trees and resolution 
complexity (e.g. \cite{beame:dp}, pp. 1) it yields a resolution proof 
of the unsatisfiability of $\Phi$
having size at most $2^{k\cdot \alpha \cdot n+1}$.   

\end{proof}
\qed

\begin{definition} 
For a formula $F$ define $
c^{*}(F)= \max\{ \frac{|Constraints(G)|}{|Var(G)|}: \emptyset \neq G 
\subseteq F\}$. 
\end{definition} 

The next result gives a sufficient condition for a generalized 
satisfiability problem 
to have a first-order phase transition. 

\begin{theorem} \label{sufficient:first-order} 
Let $C$ be a set of constraints such that $SAT({\cal C})$ has a sharp 
threshold. If 
there exists $\epsilon > 0$ such that for every minimally unsatisfiable 
formula $F$ it holds 
that 
\[
c^{*}(F) > \frac{1+\epsilon}{k-1}
\] 
then for every ${\cal P}$ with $supp({\cal P})=C$  

$SAT({\cal P})$ has a first-order phase transition.  

\end{theorem} 
\begin{proof} 

We first recall the following concept from \cite{chvatal:szemeredi:resolution}: 

\begin{definition} 
Let $x,y>0$. A $k$-uniform hypergraph with $n$ vertices is {\em ($x$,$y$)-sparse} if every set of $s\leq xn$ vertices contains at most $ys$ edges.  
\end{definition} 

We also recall Lemma 1 from the same paper.  
\begin{lemma}\label{sparsity:hypergraph}  
Let $k,c>0$ and $y$ be such that $(k-1)y>1$. Then w.h.p. the random $k$-uniform hypegraph with $n$ vertices and $cn$ edges is $(x,y)$-sparse, where 
\begin{eqnarray*}
\epsilon = y-1/(k-1), \\
x = (\frac{1}{2e}(\frac{y}{ce})^{y})^{1/\epsilon}, \\
\end{eqnarray*} 
\end{lemma} 

The critical observation is that the existence of a minimally 
unsatisfiable formula of size $xn$ and with $c^{*}(F) > \frac{1+\epsilon}{k-1}$ 
implies that the 
$k$-uniform hypergraph associated to the given formula is {\em  not}  
$(x,y)$-sparse, for $y= 
\frac{\epsilon}{k-1}$. 

But, according to Lemma~\ref{sparsity:hypergraph}, w.h.p. a random $k$-uniform hypergraph with $cn$ edges is 
$(x_{0},y)$ 
sparse, for $x_{0}=(\frac{1}{2e}(\frac{y}{ce})^{y})^{1/\epsilon}$ (a 
dirrect application of Lemma 1 in their paper). We infer that any formula 
with less than $x_{0}\cdot n /K$ 
constraints is satisfiable, therefore the same is true for any formula 
with $x_{0}\cdot n/K$ clauses picked up from the clausal representation 
of constraints in $\Phi$.  

The second condition (expansion of the formula hypergraph) can be 
proved similarly. 

\end{proof} 
\qedbox

One can give an explicitly defined class of satisfiability 
problems for which the previous result applies: 

\begin{theorem}\label{implicates:first-order}
Let ${\cal P}$ be such that $SAT({\cal P})$ has a 
sharp 
threshold. If  {\em no} clause $C\in {\cal C}=supp({\cal P})$ has an 
implicate of 
length 
at most 2 then 
\begin{enumerate}
\item for every minimally unsatisfiable formula $F$
\[
c^{*}(F)\geq \frac{2}{2k-3}. 
\]
Therefore $SAT({\cal P})$ satisfies the conditions of the previous 
theorem, i.e. it has a first-order 
phase transition.  
\item Moreover $SAT({\cal P})$ also has $2^{\Omega(n)}$ 
resolution complexity\footnote{this result subsumes some of the recent results 
in \cite{mitchell:cp02}}.    
\end{enumerate} 
\end{theorem} 

\begin{proof}

\begin{enumerate}
\item 
For any real $r \geq 1$, formula $F$ and set of clauses $G\subseteq F$, 
 define the {\em $r$-deficiency of $G$}, $\delta_{r}(G)= 
r|Clauses(G)|-|Vars(G)|$. 

Also define 
\begin{equation} \label{max}
\delta^{*}_{r}(F)= \max\{\delta_{r}(G): \emptyset \neq G \subseteq F\}
\end{equation} 

We claim that for any minimally unsatisfiable $F$, 
$\delta^{*}_{2k-3}(F)\geq 0$. Indeed, 
assume  
this was not true. Then there exists such $F$ such that: 
\begin{equation}\label{deff}  
\delta_{2k-3}(G)\leq 
-1\mbox{ for all }\emptyset \neq G\subseteq F. 
\end{equation} 
\begin{proposition} \label{1-transversal}
Let $F$ be a formula for which condition~\ref{deff} holds. Then there 
exists 
an ordering $C_{1}, \ldots, C_{|F|}$ of constraints in $F$ 
such that each constraint $C_{i}$ contains 
at least $k-2$ variables that appear in {\em no} $C_{j}$, $j<i$. 
\end{proposition} 

\begin{proof}
Denote by $v_{i}$ the number of variables that appear in {\em exactly} $i$ constraints of $F$. We have 
\[
\sum_{i\geq 1} i\cdot v_{i} = k\cdot |Constraints(F)|.
\]

therefore $2|Var(F)|-v_{1}\leq k\cdot |Constraints(F)|$. This can be 
rewritten as $v_{1}\geq 2|Var(F)|-k|Constraints(F)|> |Constraints(F)|\cdot 
(2k-3 - k)= (k-3)\cdot |Constraints(F)|$ (we have used the upper bound on $c^{*}(F)$. Therefore there exists at least one constraint in $F$  with at least $k-2$ variables that are free in $F$. We set $C_{|F|}=C$ and 
apply this argument recursively to $F\setminus C$.  
\end{proof} 
\qed

Call the $k-2$ new variables of $C_{i}$ {\em free in $C_{i}$}. Call the 
other 
two variables {\em bound in $C_{i}$}. Let us show now that $F$ cannot 
be 
minimally unsatisfiable. Construct a satisfying assignment for $F$ 
incrementally: Consider constraint $C_{j}$. At most two of the 
variables in 
$C_{j}$ are bound for $C_{j}$. Since $C$ has no implicates of size at 
most two, 
one can set the remaining variables in a way that satisfies $C_{j}$. 
This yields a satisfying 
assignment for $F$, a contradiction with our assumption that $F$ was 
minimally unsatisfiable. 

Therefore $\delta^{*}_{2k-3}(F)\geq 0$, a 
statement equivalent to our conclusion.  

\item To prove the resolution complexity lower bound we use the size-width 
connection for resolution complexity obtained in 
\cite{ben-sasson:resolution:width}: 
we prove that there exists $\eta >0$ such that w.h.p. random instances 
of $SAT({\cal P})$ having constraint density $c$ have resolution 
width at least $\eta \cdot n$. 

We use the same strategy as in \cite{ben-sasson:resolution:width} 
\begin{enumerate}
\item prove that w.h.p. minimally unsatisfiable subformulas are ``large''. 
\item prove that any clause implied by a satisfiable formula of ``intermediate'' size will contain ``many'' literals.   
\end{enumerate} 

Indeed, define for a unsatisfiable formula $\Phi$ and (possibly empty) clause 
$C$
\[
\mu(C)=\min\{|\Xi|: \Xi\subseteq  \Phi, \Xi \models C\}. 
\]

\begin{claim}
There exists $\eta_{1}>0$ such that for any $c>0$, w.h.p. 
$\mu(\Box)\geq \eta_{1}\cdot n$ (where  $\Phi$ is a random instance 
of $SAT({\cal P})$ having constraint density $c$). 
\end{claim} 

\begin{proof}
In the proof of Theorem~\ref{sufficient:first-order} we have shown that there 
exists $\eta_{0}>0$ such that w.h.p. any unsatisfiable subformula of a given 
formula has at least $\eta_{0} \cdot n$ constraints. Therefore {\em any} 
formula made of {\em clauses} in the CNF-representation of constraints in 
$\Phi$, and which has less than $\eta_{0} \cdot n$ clauses is satisfiable, 
and the claim follows, by taking $\eta_{1} = \eta_{0}$. 
\end{proof} 
\qed 

The only (slightly) nontrivial step of the proof, which critically uses the 
fact that constraints in ${\cal P}$ do not have implicates of length at most 
two, is to prove that clause implicates of subformulas of ``medium'' size 
have ``many'' variables. Formally: 

\begin{claim}\label{expansion} 
There exists $d>0$ and $\eta_{2}>0$ such that w.h.p., for every clause 
$C$ such that $d/2\cdot n <\mu(C)<=dn$, $|C|\geq \eta_{2}\cdot n$.   
\end{claim} 

\begin{proof} 
Take $0<\epsilon$. It is easy to see that if $c^{*}(F)<\frac{2}{2k-3+\epsilon}$ then w.h.p. for every subformula $G$ of $F$, at least $\frac{\epsilon}{3} \cdot |Constraints(G)|$ have at least $k-2$ private variables: Indeed, since 
$c^{*}(G)<\frac{2}{2k-3+\epsilon}$, by a reasoning similar to the one we made 
previously $v_{1}(G)\geq (k-3+\epsilon)|Constraints(G)|$. Since constraints in $G$ have arity $k$, at least $\epsilon/3 \cdot |Constraints(G)|$ have at 
least $k-2$ ``private'' variables. 

Choose $y=\frac{2}{2k-3+\epsilon}$ in Lemma~\ref{sparsity:hypergraph} for $\epsilon>0$ a small 
enough constant. Since the problem has a sharp threshold in the region where 
the number of clauses is linear, 
\[
d=\inf\{x(y,c): c>= c_{SAT({\cal P})}\} >0.  
\]

W.h.p. all subformulas of $\Phi$ having size less than  
$d/k \cdot n$ have a formula hypergraph that is $(x,y)$-sparse, 
therefore fall under the scope of the previous argument. 

Let $\Xi$ be a subformula of $\Phi$, having minimal size, such that 
$\Xi \models C$. We claim: 

\begin{claim}
For every clause $P$ of $\Xi$ with $k-2$ ``private'' variables, (i.e. one that does not appear in any other clause), at least one of these ``private'' variables appears in $C$. 
\end{claim} 

Indeed, supppose there exists a clause $D$ of $\Xi$ such that none of its 
private variables appears in $C$. 

Because of the minimality of $\Xi$ there exists an assignment $F$ that satisfies $\Xi \setminus \{D\}$ but does not satisfy $D$ or $C$. Since $D$ has no implicates 
of size two, there exists an assignment $G$, that differs from $F$ only on 
the private variables of $D$, that satisfies $\Xi$. But since $C$ does not 
contain any of the private variables of $D$, $F$ coincides with $G$ on variables in $C$. The conclusion is that $G$ does not satisfy $C$, which contradicts 
the fact that $\Xi\models C$. 

The proof of Claim~\ref{expansion} (and of item 2. of Theorem~\ref{implicates:first-order}) follows: since for any clause $K$ of one of the original constraints $\mu(K)=1$, since $\mu(\Box)>\eta_{1}\cdot n$ and since w.l.o.g. $0<d<\eta_{1}$ (otherwise replace $d$ with the smaller value) 
there exists a clause $C$ such that 
\begin{equation}\label{intermediate}
\mu(C)\in [d/2k \cdot n, d/k \cdot n].
\end{equation}

Indeed, 
 let $C^{\prime}$ be a clause in the resolution refutation of $\Phi$ minimal with 
the property that $\mu(C^{\prime})> dn$. Then at least one clause $C$, of the 
two involved in deriving $C^{\prime}$ satisfies equation~\ref{intermediate}.  

By the previous claim it 
$C$ contains at least one ``private'' variable from each clause of $\Xi$. Therefore $|C|\geq \eta_{2}\cdot n$, with $\eta_{2}=d/2k\cdot \epsilon$. 
 
\end{proof}

\end{enumerate} 
   
\end{proof} 
\qedbox

It is instructive to note that the condition in the theorem is violated 
(as expected) by random 2-SAT, as well as by random 1-in-$k$ SAT: the 
formula $C(x_{1}, x_{2}, \ldots, x_{k-1}, x_{k})\AND 
C(\overline{x_{k}}, 
x_{k+1},  \ldots , x_{2k-2}, x_{1})\AND C(\overline{x_{1}}, x_{2k-1}, 
\ldots, x_{3k-3}, \overline{x_{k}})
\AND C(x_{k}, x_{3k-2}, \ldots, x_{4k-4}, x_{1})$ (where $C$ is the 
constraint ``1-in-$k$'') is minimally unsatisfiable, but 
has clause/variable ratio $1/(k-1)$ and implicates 
$\overline{x_{1}} \OR \overline{x_{k}}$ and $x_{1}\OR x_{k}$.

It would be tempting to speculate that whenever both $x\OR y$ and 
$\overline{x}\OR \overline{y}$ are implicates of clauses in  ${\cal C}$ 
then 
$SAT({\cal P})$ has a second-order phase transitions for every 
distribution ${\cal P}$ with $supp({\cal P})= {\cal C}$. That is, 
however, {\em 
not} true, at least 
for some distributions ${\cal P}$. Consider the random $(2+p)$-SAT 
model of Monasson et al. 
\cite{2+p:nature}. In this model $p$ is a fixed real in $[0,1]$. A 
random instances of $(2+p)$-SAT with $n$ 
variables and $c\cdot n$ clauses is obtained by choosing $pcn$ random 
clauses of length 3 and $(1-p)cn$ random clauses of length 2. 
It was shown in \cite{2+p:nature} (using the nonrigorous replica 
method) that
\begin{enumerate}
\item $(2+p)$-SAT has a second-order phase transition for $0\leq p \leq 
p_{0}\sim 0.413...$. 
\item the transition becomes first-order for $p > p_{0}$. 
\item when the transition changes from second-order to first-order the 
complexity 
of a certain DPLL algorithm changes from polynomial to exponential. 
\end{enumerate}

Several rigorous results have complemented these findings. Achlioptas 
et al. \cite{2+psat:ralcom97} 
have shown that for $0\leq p \leq 0.4$ the phase transition in 
$(2+p)$-SAT only 
depends on the ``2-SAT part''. One can perhaps use the techniques 
of \cite{scaling:window:2sat} to confirm statement (i).

It is easily seen that 
$(2+p)$-SAT can be represented in Molloy's framework. 
On the other hand Achlioptas, Beame and Molloy 
\cite{achlioptas:beame:molloy} have shown that for 
those $p$ for 
which $(2+p)$-SAT does {\em not} behave like 2-SAT the resolution 
complexity of 
the problem is exponential. Using Theorem~\ref{second:order} and 
and results in \cite{achlioptas:beame:molloy} we get: 

\begin{theorem}\label{2+p-sat:first-order}
Let $p\in [0,1]$ be s.t. there exist $\epsilon >0$ and 
$c<\frac{(1-\epsilon)}{1-p}$ s.t.  
random instances of $(2+p)$-SAT with $n$ variables and $c\cdot n$ 
clauses are w.h.p. unsatisfiable. Then $(2+p)$-SAT 
has a first-order phase transition. 
\end{theorem}

It would be interesting to obtain a complete characterization of the 
order of the phase 
transition in an arbitrary problem $SAT({\cal P})$. Such a 
characterization, however, requires substantial advances: Exactly locating the 
``tricritical point'' $p_{0}$ in random $2+p$-SAT (or merely deciding 
whether it is equal or not to 0.4) is an open problem. A complete 
characterization would yield a solution to this problem as a byproduct. 

On the other hand Theorem~\ref{2+p-sat:first-order} suggests an 
interesting conjecture: whenever the location of the phase transition  
is {\em not} determined by the implicates of size at most two in the 
given formula, the phase transition in $SAT({\cal P})$ is first-order. 
Perhaps techniques in \cite{achlioptas:beame:molloy} can help settle 
this question.

\section{Discussion} 

We have shown that the existence of a first-order phase transition 
in a random satisfiability problem is often correlated with a 
$2^{\Omega(n)}$ peak in the complexity of resolution/DPLL algorithms at 
the 
transition point. 

As for the extent of the connection it is easy to see that it does not 
extend to a substantially larger class of algorithms: consider random 
$k$-XOR-SAT, the problem of testing the satisfiability of random 
systems of linear equations of size $k$ over ${\bf Z}_{2}$.  
$k$-XOR-SAT is a version of XOR-SAT, one of the polynomial time 
cases of satisfiability from Schaefer's Dichotomy Theorem 
\cite{schaefer-dich}. 
Indeed, 
it is easily solved by Gaussian elimination. But Ricci-Tersenghi et al. 
\cite{zecchina:kxorsat} have presented 
a non-rigorous argument using the replica method that supports 
the existence of a first-order phase transition, and we can show 
this formally (as a direct consequence of 
Theorem~\ref{implicates:first-order}):   

\begin{proposition}
Random $k$-XOR-SAT, $k\geq 3$, has a first-order phase transition. 
\end{proposition} 

To sum up: {\bf the intuitive argument states that a first-order phase 
transition correlates with a $2^{\Omega(n)}$ lower bound of the 
complexity 
of DPLL algorithms at the transition. This is true in many cases, and 
the underlying reason is that the two phenomena (the jump in the 
order parameter and the resolution complexity lower bound) have 
common causes. However, at least for satisfiability problems, 
this connection does not extend substantially beyond the class of 
DPLL algorithms.} 

\section*{Acknowledgments} 

I thank Madhav Marathe, Anil Kumar and Cris Moore for useful comments. 
In particular Cris made the observation that led to the realization 
that 
my previous results implied Theorem~\ref{second:order}. 

This work has been supported by the Department of Energy under contract 
W-705-ENG-36.

%\bibliography{/u1/gistrate/bib/bibtheory}
\bibliography{bibtheory}

\end{document}